\documentclass[12pt]{article}

\global\arraycolsep=1pt
\usepackage{amsmath}
\usepackage{amssymb}
\usepackage {pstricks}
\usepackage{pstcol, pst-node}
\usepackage{color}
\newcommand{\beq}{\begin{equation}}
\newcommand{\eeq}{\end{equation}}
\newcommand{\beqa}{\begin{eqnarray}}
\newcommand{\eeqa}{\end{eqnarray}}
\newcommand{\CR}{\nonumber \\}
\newcommand{\m}{\mu}
\newcommand{\n}{\nu}
\renewcommand{\k}{\kappa}
\newcommand{\ld}{\lambda}

\renewcommand{\theequation}{\thesection.\arabic{equation}}
\renewcommand{\thefootnote}{\fnsymbol{footnote}}

\newcommand{\trace}{{\rm Tr}}

\begin{document}

\begin{titlepage}
\begin{flushright}
{February, 2006} \\
{\tt hep-th/0602179} 
\end{flushright}
\vspace{0.5cm}
\begin{center}
{\Large \bf
Universal Character and Large $N$ Factorization 
in Topological Gauge/String Theory}
\vskip3.0cm
{\large Hiroaki Kanno}
\vskip 1.0em
{\it 
Graduate School of Mathematics \\
Nagoya University, Nagoya, 464-8602, Japan}
\end{center}
\vskip2.5cm


\begin{abstract}
We establish a formula of the large $N$ factorization of the modular $S$-matrix 
for the coupled representations in $U(N)$ Chern-Simons theory.
The formula was proposed by Aganagic, Neitzke and Vafa, based on
computations involving the conifold transition. We present a more
rigorous proof that relies on the universal character for rational
representations and an expression of the modular $S$-matrix 
in terms of the specialization of characters. 

\end{abstract}
\end{titlepage}


\renewcommand{\thefootnote}{\arabic{footnote}}
\setcounter{footnote}{0}


\section{Introduction}
\setcounter{equation}{0}

In \cite{OSV} a remarkable relation
\beq
Z_{BPS} \simeq |Z_{top}|^2~, \label{OSVrel}
\eeq
was proposed. The proposal (\ref{OSVrel}) relates the partition function $Z_{BPS}$
of counting BPS black holes formed by $D$-branes and topological string amplitudes $Z_{top}$.
While several attempts have been made at making (\ref{OSVrel}) more precise, 
the conjecture has been tested for non-compact Calabi-Yau 3-folds in 
\cite{Vafa},\cite{AOSV},\cite{ANV},\cite{AJS}, where BPS microstates arise from $D4$-branes wrapping over
a line bundle ${\cal O}(-n) \to \Sigma$ on a Riemann surface $\Sigma$. The (electric) charges of black hole
are given by the numbers of $D0$ and $D2$ branes bound to $D4$ branes. 
When we have $N$ $D4$ branes, the partition function $Z_{BPS}$ is computed from
$U(N)$ topological Yang-Mills theory on the world volume of $D4$ branes and it has been shown
that it is reduced to the $q$-deformed Yang-Mills theory on $\Sigma$ \cite{AOSV}.
The large $N$ factorization of two dimensional Yang-Mills theory and its interpretation by a closed
string theory were given in \cite{GT1},\cite{GT2}. The group theory of coupled representation
labeled by a pair of Young diagrams was employed there. In the $q$-deformed case, 
the key to the large $N$ factorization of the partition function is the following formula given in \cite{ANV};
\beqa
q^{\rho_{N}^2 + \frac{N}{24}}S_{\cal P Q} (q,N)
&=& M(q) \eta(q)^N q^{\frac{1}{2} ( \k_{Q_+} + \k_{Q_-} )}
        \cdot q^{\frac{N}{2}(|P_+| + |P_-| +|Q_+| + |Q_-|) } \CR
       & & \times \sum_R (-1)^{|R|} q^{-N|R|} C_{P_+ Q^t_+ R}(q) C_{P_- Q^t_- R^t}(q)~, \label{intro}
\eeqa
for an extension  $S_{\cal P Q}$ of the modular matrix of the $U(N)$
Chern-Simons theory to the coupled representations ${\cal P}, {\cal Q}$.
In view of  the conjecture (\ref{OSVrel}), it is important that 
in (\ref{intro}) we have the topological vertex $C_{PQR}(q)$, which is
a building block of all genus topological string amplitudes 
on local toric Calabi-Yau 3-folds \cite{AMV},\cite{AKMV},\cite{LLLZ}.
On the other hand, the modular $S$-matrix $S_{\cal P Q}$ appears as
a building block for $Z_{BPS}$ \cite{AOSV},\cite{AJS}.
For example, for $D4$ branes wrapping over a non-compact four cycle ${\cal O}(-p) \to {\bf P}^1$, 
the BPS partition function is
\beq
Z({\bf P}^1, p) = \sum_{\cal R} \left( S_{{\cal R} \bullet} \right)^2 
q^{\frac{p}{2}C_2({\cal R})} e^{i\theta C_1({\cal R})}~,
\eeq
where $C_1({\cal R})$ and $C_2({\cal R})$ are the Casimir of irreducible $U(N)$ representation
$\cal R$ and $\theta$ is the $\theta$ angle of the $q$-deformed $YM_2$. In this paper
we will use $\bullet$ for the trivial representation. 
We note that $S_{{\cal R} \bullet}$ is related to the quantum dimension 
by $S_{{\cal R} \bullet} = S_{\bullet \bullet} \cdot {\rm dim}_q {\cal R}$.
Up to the normalization associated with $S_{\bullet \bullet}$, 
the formula (\ref{intro}) gives
\beq
S_{{\cal R} \bullet} (q,N) =  q^{\frac{N}{2}(|R_+| + |R_-|)} 
\cdot \sum_Q (-1)^{|Q|} q^{-N|Q|} C_{R_+ \bullet Q}(q) C_{R_- \bullet Q^t}(q)~.
\eeq
Since only the trivial representation $Q=\bullet$ survives in the large $N$ limit, 
we obtain\footnote{We use $C_1({\cal R}) = C_1(R_+) - C_1(R_-)$ and
$C_2({\cal R}) = C_2(R_+) + C_2(R_-) + O(1/N)$.}
\beq
Z({\bf P}^1, p) \sim Z_{top}^+ \cdot Z_{top}^-~, \qquad
Z_{top}^\pm := \sum_{R_\pm} \left( C_{R_\pm \bullet \bullet}(q) \right)^2 
q^{\frac{p}{2} C_2({R_\pm})} e^{\pm i\theta C_1(R_\pm)}~.
\eeq
The factor $Z_{top}^\pm$ is the topological string amplitude
on the local toric Calabi-Yau manifold : ${\cal O}(p-2) \oplus
{\cal O}(-p) \to {\bf P}^1$ with an appropriate (complexified) K\"ahler parameter
\footnote{See also \cite{CCGPSS} for the relation of topological string amplitudes
to the $q$-deformed $2D$ Yang-Mills theory.}.
Thus we find the topological string amplitudes in
the factorization of $Z_{BPS}$ at large $N$.

In \cite{ANV} the validity of the formula (\ref{intro}) has been argued by comparing the open/closed 
topological string amplitudes related by the conifold transition. 
The main purpose of this article is to give a more rigorous proof based on the universal character
of the coupled representation labeled by a pair of partitions $(\ld, \m)$ or Young diagrams.
The Schur function $s_\ld(x)$ is nothing but the universal character of irreducible (polynomial) 
representation corresponding to a single partition $\ld$. 
The coupled representation used in \cite{ANV} is what is 
called {\it rational} representation in mathematics literature
and the universal character  $s_{[\ld, \m]}(x,y)$ of the rational representation 
is a generalization of the Schur function defined, for example, in \cite{Koi}.
This article is organized as follows; In section 2 we introduce the universal character
for rational representation following \cite{Koi}. We will see that $s_{[\ld, \m]}(x,y)$ 
is expanded in terms of (skew) Schur functions. 
In section 3 we review how the Hopf link invariants, or
the normalized modular $S$-matrices of the Chern-Simons-WZW theory, are
expressed in terms of specialization of Schur functions \cite{ML}. These facts are basic
ingredients in our proof of (\ref{intro}), since it is known that the topological 
vertex has an expression in terms of specialization of skew Schur functions \cite{ORV}. 
Finally we prove (\ref{intro}) in section 4 by using the formulas for
the specialization of skew Schur function. In might be interesting to investigate
some applications of the formula  (\ref{intro}) to the large $N$ limit of the discrete matrix model
and the generalized two-dimensional Yang-Mills theory \cite{MO}.


\section{Universal character for rational representation}
\setcounter{equation}{0}

A finite dimensional representation $X \to \rho(X)$ of 
$GL(N, {\mathbb C})$\footnote{$U(N)$ and $SU(N)$ are the compact real forms 
of $GL(N, {\mathbb C})$ and $SL(N, {\mathbb C})$, respectively.} is
called polynomial or rational, if the matrix components of $\rho(X)$ are 
polynomial or rational functions of the components of $X$, respectively. 
For example the determinant $\rho(X) = {\rm det}~X$ is a one-dimensional polynomial 
representation. The Cartan subalgebra of $GL(N, {\mathbb C})$ is the set of 
diagonal matrices ${\mathfrak h}_N := \{ X ; X={\rm diag.}(x_1, x_2, \cdots, x_N) \}$. 
The weight lattice is generated by $\epsilon_i : {\mathfrak h}_N  \to {\mathbb C}, \epsilon_i(X):= x_i$.
It is well-known that the irreducible polynomial representations of 
$GL(N, {\mathbb C})$ are labeled by partitions $\m : \m_1 \geq
\m_2 \geq \cdots \geq \m_{\ell(\m)} > 0$  of length $\ell(\m) \leq N$. 
The highest weight of the representation $\rho_\m$ labeled by $\m$ is 
$\mu_1 \epsilon_1 + \mu_2 \epsilon_2 + \cdots + \mu_N \epsilon_N$ and 
the character is known as the Schur polynomial 
$\trace_{\rho_\m} X= s_\m(x_1, \cdots, x_N)$, where
$(x_1, \cdots, x_N)$ are eigenvalues of $X$. For a fixed partition
$\m$, the projective limit $s_\m(x) = \varprojlim s_\m(x_1, \cdots, x_N)$ 
has the meaning in the ring of symmetric functions $\Lambda_x$, which is the projective
limit of the algebra of symmetric polynomials in $N$ variables.
In this sense the Schur function $s_\m(x) \in \Lambda_x$ is
sometimes called universal character of polynomial representation.

The irreducible rational representations of $GL(N, {\mathbb C})$ were
classified by Schur. He showed any rational representation is of the
form $X \to ({\rm det}~X)^r \rho(X)$, where $r \in {\mathbb Z}$ and
$\rho$ is a polynomial representation. That is, the denominator of
rational representation is a power of the determinant. We may identify
the integer $r$ with the $U(1)$ charge of the rational representation.
The highest weight of the determinant representation is $\sum_{i=1}^N \epsilon_i$.
Thus a complete set of inequivalent 
rational representations of $GL(N, {\mathbb C})$ is indexed by
$N$-tuples $\ld \in {\mathbb Z}^N : \ld_1 \geq \ld_2 \geq \cdots \geq \ld_N$ 
and the highest weight is given by $\sum_{i=1}^N \ld_i \epsilon_i$. 
The polynomial representations are characterized by the condition $\ld_N \geq 0$.
Separating the (strictly) positive and negative parts of $\ld$, we can also label them
in terms of a pair of partitions $(\ld_+, \ld_-)$ with $\ell(\ld_+) + \ell(\ld_-)
\leq N$ so that $(\ld_1, \ld_2, \cdots, \ld_N)=
(\ld_{+,1}, \cdots, \ld_{+,\ell(\ld_+)}, 0,\cdots, 0,-\ld_{-,\ell(\ld_-)}, \cdots, -\ld_{-,1})$.
In $SU(N)$ gauge theory the difference of polynomial representations and
rational representations is irrelevant, since the determinant representation
is trivial. However, for $U(N)$ gauge theory the difference is important.
In fact Gross and Taylor employed the rational representation,
which they called composite representation, to work out the large $N$ factorization 
of the partition function of two-dimensional Yang-Mills theory \cite{GT1},\cite{GT2}.


\begin{center}
\begin{pspicture}(-4,0)(10,5) 
\psline(-1.5,0)(0,0)\psline(-1.5,0.5)(-1,0.5)\psline(-1,1)(0,1)
\psline(0,2)(0.5,2)\psline(0.5,3)(1,3)\psline(1,3.5)(1.5,3.5)
\psline(1.5,4)(2.5,4)\psline(0,4.5)(2.5,4.5)
\psline(-1.5,0)(-1.5,0.5)\psline(-1,0.5)(-1,1)\psline(0,0)(0,1.2)
\psline(0,1.2)(0,1.8)
\psline(0,1.8)(0,4.5)\psline(0.5,2)(0.5,3)\psline(1,3)(1,3.5)
\psline(1.5,3.5)(1.5,4)\psline(2.5,4)(2.5,4.5)
\rput(0.7,3.8){$\ld_+$} \rput(-0.5, 0.5){$\ld_-$}
\rput(3.7,2){\large $\simeq$}
\psline(5.5,0.5)(6,0.5)\psline(6,1)(7,1)
\psline(7,2)(7.5,2)\psline(7.5,3)(8,3)\psline(8,3.5)(8.5,3.5)
\psline(8.5,4)(9.5,4)\psline(7,4.5)(9.5,4.5)
\psline(6,0.5)(6,1)\psline(7,1)(7,2)
\psline(7.5,2)(7.5,3)\psline(8,3)(8,3.5)
\psline(8.5,3.5)(8.5,4)\psline(9.5,4)(9.5,4.5)
\psline(5.5,0.5)(5.5,4.5)\psline(7,4.5)(5.5,4.5)
\psframe[linestyle=none, fillstyle=hlines, linewidth=0.4pt](5.5,0)(7,4.5)
\end{pspicture}

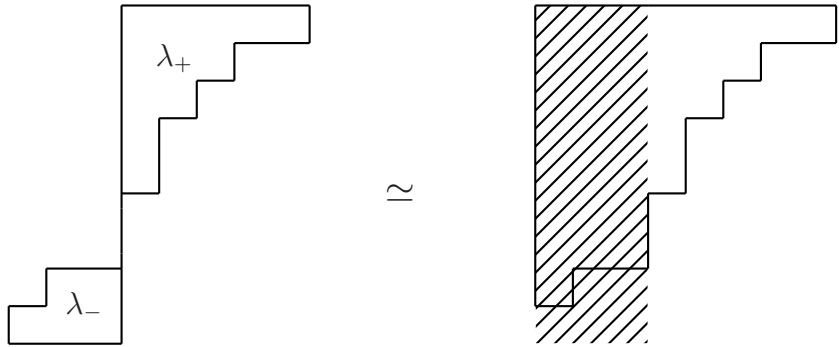
Figure 1: Rational representation as a tensor product of the determinant (the 
shaded rectangle) and 
a polynomial representation. 
\end{center}


Let $\ld_\pm$ be a partition of $|\ld_\pm| := \sum_{i=1}^{\ell(\ld_\pm)} \ld_{\pm,i}$. 
Then it is known that the rational representation ${\cal R}$ 
labeled by $(\ld_+, \ld_-)$ occurs in the
tensor product of $|\ld_+|$ copies of the defining representation ${\mathbb C}^N$ and
$|\ld_-|$ copies of the contragredient representation $({\mathbb C}^N)^*$;
\beq
T_{|\ld_+|, |\ld_-|} := \left( \stackrel{|\ld_+|}{\otimes} {\mathbb C}^N \right)
\bigotimes \left( \stackrel{|\ld_-| }{\otimes}({\mathbb C}^N)^* \right) ~.
\eeq
The universal character of the rational representation has been introduced in \cite{Koi}. 
Let us denote the universal character of the rational representation ${\cal R}$ by
$s_{[\ld_+, \ld_-]}(x,y)$, where the variables $x$ and $y$ are associated with
the defining representation ${\mathbb C}^N$ and the contragredient representation
$({\mathbb C}^N)^*$, respectively. Consequently, $s_{[\ld_+, \ld_-]}(x,y)$ is in the tensor
product of universal character rings $\Lambda_x \otimes \Lambda_y$ and
it can be expanded in terms of the products of Schur functions. Let us introduce
the Littlewood-Richardson coefficients $c_{\eta, \n}^\m$ defined by the relation;
\beq
s_\eta(x) s_\n(x) = \sum_\m c_{\eta, \n}^\m s_\m(x)~.
\eeq
The coefficients $c_{\eta, \n}^\m$ are non negative integers.
In \cite{Koi} the expansion
\beq
s_{[\ld_+, \ld_-]}(x,y) = \sum_{\tau, \eta, \n}  (-1)^{|\tau|} c_{\eta, \tau}^{\ld_+} 
c_{\n, \tau^t}^{\ld_-} s_\eta(x) s_\n (y)~, \label{expand}
\eeq
and its inverse relation
\beq
s_\eta(x) s_\n (y)  = \sum_{\tau, \ld_+, \ld_-}  c_{\ld_+, \tau}^{\eta} 
c_{\ld_-,\tau}^{\n} s_{[\ld_+,\ld_-]}(x,y)~,
\eeq
were proved. The partition $\tau^t$ in (\ref{expand}) is defined by the transpose of the
Young diagram. Using the skew Schur function defined by
\beq
s_{\m/\n} (x) = \sum_\tau c_{\tau, \n}^\m s_{\tau} (x)~,
\eeq
we can write the expansion (\ref{expand}) as
\beq
s_{[\ld_+, \ld_-]}(x,y) = \sum_{\tau}  (-1)^{|\tau|} 
s_{\ld_+/\tau} (x) s_{\ld_-/\tau^t} (y)~.
\eeq
Finally we note that there is the embedding $GL(N, {\mathbb C}) \ni g \mapsto (g, (g^t)^{-1}) 
\in GL(N, {\mathbb C}) \times GL(N, {\mathbb C})$, which is a group
homomorphism. We can use the induced homomorphism $\Lambda_x \otimes
\Lambda_y \to \Lambda_x$ obtained by putting $y= x^{-1}$ to send the above formulas 
to relations in the universal character ring $\Lambda_x$ of $GL(N, {\mathbb C})$.

\section{Hopf link invariants and specialization of Schur function}
\setcounter{equation}{0}

Both the elementary symmetric functions $e_1(x), \cdots, e_n(x), \cdots$ and 
the complete symmetric functions $h_1(x), \cdots, h_n(x), \cdots$ are
${\mathbb Z}$-basis of the ring $\Lambda_x$ of the symmetric functions.
They are defined by the following generating functions;
\beqa
E(t, x) &:=& 1 + \sum_k e_k(x) t^k = \prod_{i=1}^\infty \left( 1 + x_i t \right)~, \label{egen} \\
H(t, x) &:=& 1 + \sum_k h_k(x) t^k = \prod_{i=1}^\infty \left( 1 - x_i t \right)^{-1}~,
\eeqa
that satisfy $E(t, x) H(-t,x) =1$. We note that the power sum functions
$p_1(x), \cdots, p_n(x), \cdots$ are not ${\mathbb Z}$-basis, but 
${\mathbb Q}$-basis of the symmetric functions. The generating function is
\beq
P(t, x) := \sum_k p_k(x) t^{k-1} = \frac{d}{dt} \log H(t, x)~.
\eeq
Any symmetric function $f \in \Lambda_x$ can be written as a polynomial
$f(e_1, \cdots, e_n, \cdots)$ in the elementary symmetric functions. For example,
the Jacobi-Trudi formula gives the Schur function in terms of $\{ e_i(x) \}$;
\beq
s_\m (x) = {\rm det}~\left( e_{\m_i^t - i +j} (x)\right) ~.
\eeq

We can define a specialization of the elementary symmetric functions 
by taking the generating function $E(t) = 1 + \sum_{n=1}^\infty e_n t^n$, 
to be any formal power series with the leading coefficient $1$. 
We denote $f(e_1, \cdots, e_n, \cdots)$ as $f(E(t))$, which gives a specialization of 
the symmetric function $f$. A basic example of such a specialization of symmetric functions 
is given in \cite{Mac}  (Examples I-2.5 and I-3.3).
When we take the generating functions 
\beq
H(t) = \prod_{i=0}^\infty \frac{1- b q^i t} {1- a q^i t}~, \qquad
E(t) = \prod_{i=0}^\infty \frac{1+ a q^i t}{1+ b q^i t}~,
\eeq
we have
\beq
h_n(q) =  \prod_{i=1}^n \frac{a- bq^{i-1}}{1 - q^i}~, \qquad
e_n(q) = \prod_{i=1}^n \frac{a q^{i-1} - b}{1-q^i}~.
\eeq
In this specialization the power sum function is 
\beq
p_n(q) = \frac{a^n - b^n}{1 - q^n}~,
\eeq
and the Schur function is given by
\beq
s_\m(q) = q^{n(\m)} \prod_{(i,j)\in \m} \frac{a - b q^{c(i,j)}}{1 - q^{h(i,j)}}~,
\eeq
where $n(\m)$ is
\beq
n(\m) = - \frac{1}{2} \sum_{(i,j)\in \m} (1 + c(i,j) - h(i,j))~.
\eeq
We have introduced the standard notation for the content $c(i,j):= j-i$ and
the hook length $h(i,j):= \m_i + \m_j^t -i -j +1$ of the box at $(i,j)$ in the Young diagram.

Morton and Lukac used this kind of specialization of Schur functions
to express the quantum dimension and Hopf link invariants \cite{ML}. 
For example, when $a= \ld^{-1}, b=1$, 
\beqa
s_\m(q) &=& q^{\frac{1}{2} \sum_{(i,j)\in \m} (h(i,j) - c(i,j) -1)} 
\prod_{(i,j)\in \m} \frac{\ld^{-1} -  q^{c(i,j)}}{1 - q^{h(i,j)}}~, \CR
&=& \left( q \ld \right)^{-\frac{|\m|}{2}} \prod_{(i,j)\in \m} \frac{[c(i,j)]_\ld}{[h(i,j)]}~. \label{qdim}
\eeqa
We find the quantum dimension ${\rm dim}_q~R=\prod_{(i,j)\in \m} \frac{[c(i,j)]_\ld}{[h(i,j)]}$ 
of the representation $R$ defined by the partition $\m$. In the $SU(N)$ Chern-Simons theory 
at level $k$, $q=\exp\left( \frac{2\pi i}{N+k} \right)$ and $\ld = q^N$. Thus we have
\beq
W_\m (q, \ld) := {\rm dim}_q~R = \ld^{\frac{|\m|}{2}} s_\m \left( 
E_{\bullet} (t; q, \ld) \right)~,
\eeq
where
\beq
E_\bullet (t) := 1 + \sum_{n=1}^\infty (q^{-\frac{1}{2}}t)^n 
\prod_{i=1}^n \frac{1 - \ld^{-1} q^{i-1}}{q^i -1}
= \prod_{i=1}^{\infty}\frac{ 1+ \ld^{-1} q^{i-\frac{1}{2}}t}
{1+ q^{i-\frac{1}{2}} t}~. \label{gen}
\eeq
We have absorbed the overall factor $q^{-\frac{|\m|}{2}}$ in (\ref{qdim})
by the shift of the exponent $q^{i}\to q^{i - \frac{1}{2}}$.
In this paper we will take the convention\footnote{This is the 
same as \cite{AKMV}, but different from \cite{AOSV},\cite{ANV} and \cite{AJS}, 
where $q = \exp (-g_s)$.} $q = \exp (g_s)$ with
$g_s$ being string coupling (the parameter of genus expansion
of topological string theory). Thus the relevant region of $q$ in our convention is
$|q| > 1$. Since the expression in (\ref{gen}) is that for
$|q| < 1$, we make an analytic continuation to the region
$|q| > 1$  \cite{IK},\cite{Zhou}, to obtain
\beq
E_\bullet (t) = \prod_{i=1}^{\infty} \frac{1+ q^{-i+\frac{1}{2}} t}
{ 1+ \ld^{-1} q^{-i+\frac{1}{2}}t}~.
\eeq
We note that $SU(N)$ specialization  $\ld= q^N$ gives
\beq
E_\bullet^{(N)} (t) = \prod_{i=1}^{\infty} 
\frac{1+ q^{-i+\frac{1}{2}} t}{ 1+ q^{-N-i+\frac{1}{2}}t}
=  \prod_{i=1}^{N} \left( 1+ q^{-i+\frac{1}{2}} t \right)~,
\eeq
which is the same as the generating function (\ref{egen})
for the specialization $x_i = q^{-i + \frac{1}{2}}~(1 \leq i \leq N), x_j=0~(N < j)$.
Similarly, the Hopf link invariants are also expressed in terms of 
an appropriate specialization of Schur function as follows; 
\beq
W_{\m\n}(q, \ld) = W_\m(q, \ld) \ld^{|\n|/2} s_\n \left(
E_\m ( t : q, \ld) \right)~,
\eeq
where
\beq
E_\m (t) := \prod_{i=1}^{\ell(\m)} \frac{1+ q^{\m_i -i+\frac{1}{2}} t}
{ 1+ q^{-i+\frac{1}{2}}t} \cdot E_\bullet (t) 
=\prod_{i=1}^{\infty} \frac{1+ q^{\m_i -i+\frac{1}{2}} t}
{1+ \ld^{-1} q^{-i+\frac{1}{2}}t}~.
\eeq
When $N \to \infty$, $\ld^{-1} = q^{-N} \to 0$ for $|q|>1$ and consequently
\beq
W_{\m\n}(q) \simeq \ld^{\frac{1}{2} (|\m|+|\n| )} s_\m(q^\rho) s_\n(q^{\m + \rho})~, \label{Hopf}
\eeq
where $q^{\rho}$ and $q^{\m+\rho}$ mean the specialization
$x_i = q^{-i +\frac{1}{2}}$ and $x_i = q^{\m_i -i + \frac{1}{2}}$,
respectively.


\section{Large $N$ factorization of the modular matrix of $U(N)$ CS theory}
\setcounter{equation}{0}


The modular S-matrix $S_{\cal P Q}$ of the $U(N)$
Chern-Simons theory for coupled representations ${\cal P}, {\cal Q}$ is
defined by
\beq
 S_{\cal P Q} (q, N) = \sum_{w \in S_N} (-1)^{w} q^{-w({\cal P} + \rho_N)\cdot({\cal Q} + \rho_N)}~, \label{modular}
\eeq
where the symmetric group $S_N$ is the Weyl group of $U(N)$ and $\rho_N$ is the Weyl vector
with the components $\frac{1}{2}(N-2i+1), i=1, \cdots, N$. Note that the same notation ${\cal P}, {\cal Q}$ is
used for the highest weight. When ${\cal P}$ and ${\cal Q}$ are
integrable representations $P, Q$ of $SU(N)$ affine Lie algebra at level $k$, (\ref{modular}) gives 
the matrix element of the $S$-transformation on the space of conformal blocks, which is
the physical Hilbert space of the Chern-Simons theory on $T^2 \times {\mathbb R}$. The Hopf link invariants
are obtained as the normalized $S$-matrix elements; $W_{PQ} = S_{PQ}/S_{\bullet\bullet}$.
The definition (\ref{modular}) is a formal extension of the modular S-matrix to rational representations of
$U(N)$. Recall that $q= \exp \left( \frac{2\pi i}{N+k} \right) = \exp (g_s)$.
In \cite{ANV} the following formula for $S_{\cal P Q}$ was claimed;
\beqa
q^{\rho_{N}^2 + \frac{N}{24}}S_{\cal P Q} (q,N)
&=& M(q) \eta(q)^N q^{\frac{1}{2} ( \k_{Q_+} + \k_{Q_-} )}
        \cdot q^{\frac{N}{2}(|P_+| + |P_-| +|Q_+| + |Q_-|) } \CR
       & & \times \sum_R (-1)^{|R|} q^{-N|R|} C_{P_+ Q^t_+ R}(q) C_{P_- Q^t_- R^t}(q)~, \label{factor}
\eeqa
where
\beq
{\cal P} = [P_+, P_-]~, \quad {\cal Q} = [Q_+, Q_-]~,
\eeq
are two coupled representations and $C_{PQR}(q)$ is the topological vertex\footnote{In
\cite{ANV}  $q:= \exp (-g_s)$, which is different from ours.}. For the representation $R$
corresponding to a partition $\m^R$, $\k_R := 2 \sum_{(i,j) \in \m^R} c(i,j)$. 
In this section we identify the representations, 
the partitions and the Young diagrams and often use the same notation for them.
The MacMahon function $M(q)$ and the eta function $\eta(q)$ defined by 
\beq
M(q) = \prod_n \frac{1}{(1 - q^n)^n}~, \quad
\eta(q) = q^{1/24}  \prod_n (1 - q^n)~,
\eeq
appear as overall factors and they are related to the normalization of the Hopf
link invariants (see Appendix).

Let us look at some special cases of the formula (\ref{factor}).
When $P_\pm = Q_\pm = \bullet$, $S_{\bullet \bullet}$ is nothing but
the partition function of $U(N)$ Chern-Simons theory on $S^3$ and (\ref{factor}) gives
\beqa
q^{\rho_{N}^2 + \frac{N}{24}} S_{\bullet \bullet} 
&=& M(q) \eta(q)^N \sum_R (-1)^{|R|} q^{-N|R|} C_{\bullet \bullet R}(q) C_{\bullet \bullet R^t}(q)~, \CR
 &=& M(q) \eta(q)^N \prod_{n=1}^\infty \left( 1 - \ld^{-1} q^{-n} \right)^n \CR
&=& M(q) \eta(q)^N \exp \left( - \sum_{n=1}^\infty \frac{e^{-tn}}
{n(q^{\frac{n}{2}}-q^{-\frac{n}{2}})^2} \right)~.
\eeqa
If we identify the 't Hooft coupling $t=Ng_s$ with the K\"ahler parameter of the resolved
conifold geometry,  we recover the basic example of gauge/geometry correspondence
of Gopakumar-Vafa \cite{GV}. When $P_- = Q_- = \bullet$, the left hand side is the usual modular matrix
of $U(N)$ theory and  (\ref{factor}) implies
\beqa
W_{PQ}(q,N) &=&  \frac{S_{PQ}(q,N)}{S_{\bullet\bullet}(q,N)}~, \CR
&=& q^{\frac{1}{2} \k_Q + \frac{N}{2}(|P| + |Q|)}  \prod_{n=1}^\infty \left( 1 - \ld^{-1} q^{-n} \right)^{-n}
 \sum_R (-1)^{|R|} q^{-N|R|} C_{PQ^t R} C_{\bullet\bullet R^t}, \label{special}
\eeqa
which has been derived in \cite{IK}.
When $N \to \infty$ only the trivial representation survives in the sum and
\beq
W_{PQ}(q)= q^{\frac{1}{2} \k_Q + \frac{N}{2}(|P| + |Q|)}  
C_{PQ^t \bullet} (q)~.
\eeq
We recover the formula (\ref{Hopf}) for $W_{PQ}(q)$.

Topological vertex is expressed in terms of the skew Schur function \cite{ORV};
\beq
C_{R_1 R_2 R_3} (q) = q^{\frac{\kappa_{R_3}}{2}} s_{R_2}(q^\rho) 
\sum_{Q} s_{R_1/Q}(q^{\mu^{R_2^t}+\rho}) s_{R_3^t/Q}(q^{\mu^{R_2}+\rho})~.
\eeq
Using the Cauchy formula
\beq
\sum_R s_{R/R_1}(x) s_{R^t/R_2}(y) = \prod_{i,j \geq 1} (1+ x_i y_j) 
\sum_Q s_{R_2^t/Q}(x) s_{R_1^t/Q^t}(y)~,
\eeq
and the relation
\beq
s_{P/Q} (q^{R+\rho}) = (-1)^{|P|-|Q|} s_{P^t/Q^t} (q^{- R^t -\rho})~,
\eeq
we can compute
\footnote{We use the cyclic symmetry of $C_{R_1 R_2 R_3}$. This type of
computation appears in confirming the flop invariance of topological string
amplitudes \cite{IK}, \cite{KM}.}
\beqa
& & q^{\frac{1}{2} ( \k_{Q_+} + \k_{Q_-} )} 
\sum_R (-1)^{|R|} q^{-N|R|} C_{R P_+ Q^t_+ }(q) C_{R^t P_- Q^t_- }(q) \CR
&=& s_{P_+}(q^\rho)s_{P_-}(q^\rho) \prod_{1\leq i,j} 
\left( 1 - \ld^{-1} q^{P^t_{+,i} + P^t_{-,j} -i -j +1}\right) \CR
& &~~~\sum_V (-\ld)^{-|V|} s_{Q_+/V} (\ld^{-1} q^{-P_- -\rho}, q^{P_+ + \rho})
s_{Q_-/V^t}(\ld^{-1} q^{-P_+ -\rho}, q^{P_- + \rho})~,
\eeqa
where the Schur function with two variables $s_R(x,y)$ is defined by
\beq
s_R(x,y) = \sum_{P,Q}  c_{PQ}^{R} s_P(x) s_Q(y) = \sum_{Q} s_{R/Q}(x) s_Q(y)~.
\eeq
The corresponding generating function of the elementary symmetric functions is
\beq
E(t) = \prod_{i=1}^\infty \left( 1 + x_i t \right) 
\prod_{j=1}^\infty  \left( 1 + y_j t \right)~.
\eeq
The cyclic symmetry of the topological vertex, 
\beq
\sum_R (-1)^{|R|} q^{-N|R|} C_{R P_+ \bullet}(q) C_{R^t P_- \bullet}(q)
=\sum_R (-1)^{|R|} q^{-N|R|} C_{P_+ \bullet R }(q) C_{P_- \bullet R^t}(q)~,
\eeq
implies
\beqa
& & s_{P_+}(q^\rho)s_{P_-}(q^\rho) \prod_{1\leq i,j} 
\left( 1 - \ld^{-1} q^{P^t_{+,i} + P^t_{-,j} -i -j +1}\right)  \CR
&=&\prod_{n=1}^\infty \left( 1 - \ld^{-1} q^{-n} \right)^n
\ld^{-|P_-|} \sum_V (-1)^{|V|} s_{P_+/V^t}(\ld^{-1} q^{-\rho}, q^\rho)
s_{P_-/V}(\ld q^{ \rho} , q^{-\rho})~.
\eeqa
Hence we finally obtain
\beqa
& & q^{\frac{1}{2} ( \k_{Q_+} + \k_{Q_-} )} 
\sum_R (-1)^{|R|} q^{-N|R|} C_{R P_+ Q^t_+ }(q) C_{R^t P_- Q^t_- }(q) \CR
&=&  \prod_{n=1}^\infty \left( 1 - \ld^{-1} q^{-n} \right)^n
\ld^{-(|P_-|+ |Q_-|)} \sum_V (-1)^{|V|} s_{P_+/V^t}(\ld^{-1} q^{-\rho}, q^\rho)
s_{P_-/V}(\ld q^{ \rho} , q^{-\rho} )\CR
& &~~~\sum_V (-1)^{|V|} s_{Q_+/V} (\ld^{-1} q^{-P_- -\rho}, q^{P_+ + \rho})
s_{Q_-/V^t}(q^{-P_+ -\rho}, \ld q^{P_- + \rho})~. \label{4vertex}
\eeqa

On the other hand, the result reviewed in section 3 formally implies an expression of 
$S_{\cal P Q}$ in terms of the specialization the Schur function $s_{\cal R}(x)$;
\beq
W_{\cal P Q}(q, N) := \frac{S_{\cal P Q}(q, N)}{S_{\bullet\bullet}(q, N)}
 = \ld^{\frac{1}{2}(|{\cal P}| + |\cal Q|)} s_{\cal P} (E_\bullet(t; q,\ld)) s_{\cal Q} (E_{\cal P} (t; q,\ld))~.
\eeq
But, we have to make it precise what $s_{\cal R}(x)$ and $E_{\cal R}(t; q,\ld)$ mean for
the rational representation $\cal R$. Firstly,
the Schur function $s_{\cal R}(x)$ should be regarded as 
the universal character of rational representation defined by the pair of partitions
$(R_+, R_-)$. The generating function $E_{\cal R}(t)$
that defines the specialization can be obtained
by looking at the Young diagram of the rational representation
${\cal R} = [R_+, R_-]$ (see Figure 2). Assuming $\ell(R_+) + \ell(R_-) \leq N$, we find
\beqa
E_{\cal R} (t) &=& \prod_{i=1}^{\ell(R_+)} \frac{1+ q^{R_{+,i} -i+\frac{1}{2}} t}
{ 1+ q^{-i+\frac{1}{2}}t} 
\prod_{j=1}^{\ell(R_-)} \frac{1+ q^{-N - R_{-,j} +j-\frac{1}{2}} t}
{ 1+ q^{-N+ j-\frac{1}{2}}t} \cdot E_\bullet (t)~, \CR
&=&\prod_{i=1}^{\infty} \frac{1+ q^{R_{+,i} -i+\frac{1}{2}} t}
{ 1+ q^{-i+\frac{1}{2}}t} \prod_{j=1}^{\infty} \frac{1+ \ld^{-1} q^{- R_{-,j} +j-\frac{1}{2}} t}
{ 1+ \ld^{-1} q^{ j-\frac{1}{2}}t} 
\prod_{k=1}^{\infty} \frac{1+ q^{-k+\frac{1}{2}} t}
{ 1+ \ld^{-1} q^{-k+\frac{1}{2}}t}~, \CR
&=& \prod_{i=1}^{\infty} \left( 1+ q^{R_{+,i} -i+\frac{1}{2}} t\right)
\prod_{j=1}^{\infty} \left( 1+ \ld^{-1} q^{- R_{-,j} +j-\frac{1}{2}} t\right)~. \label{extend}
\eeqa

\begin{center}
\begin{pspicture}(-4,0)(3,5) 
\psline(-1.5,0)(0,0)\psline(-1.5,0.5)(-1,0.5)\psline(-1,1)(0,1)
\psline(0,2)(0.5,2)\psline(0.5,3)(1,3)\psline(1,3.5)(1.5,3.5)
\psline(1.5,4)(2.5,4)\psline(0,4.5)(2.5,4.5)
\psline(-1.5,0)(-1.5,0.5)\psline(-1,0.5)(-1,1)\psline(0,0)(0,1.2)
\psline[linestyle=dashed,dash=3pt 2pt](0,1.2)(0,1.8)
\psline(0,1.8)(0,4.5)\psline(0.5,2)(0.5,3)\psline(1,3)(1,3.5)
\psline(1.5,3.5)(1.5,4)\psline(2.5,4)(2.5,4.5)
\psline[linestyle=dashed](-3,4.5)(0,4.5)
\psline[linestyle=dashed](-3,0)(-1.5,0)
\psline[arrowsize=5pt]{->}(-2,1.5)(-2,0)
\psline[arrowsize=5pt]{->}(-2,3)(-2,4.5)
\rput(-0.5,0.5){\large$R_-$} \rput(0.75,3.75){\large$R_+$}
\rput(-2,2.25){$N$}
\end{pspicture}

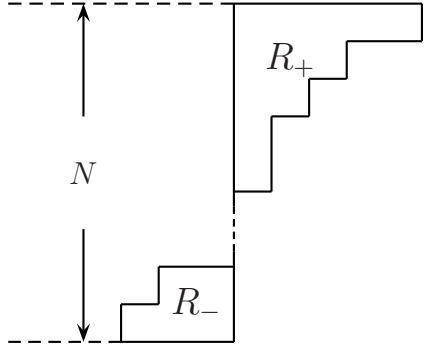
Figure 2: Extended Young diagram for rational representation labeled by $(R_+, R_-)$. 
\end{center}


\noindent
Since the universal character of the rational representation $\cal R$ is
given by
\beq
s_{\cal R}(x) = \sum_Q (-1)^{|Q|} s_{R_+/Q}(x) s_{R_-/Q^t}(x^{-1})~,
\eeq
and the specialization of $s_{\cal R}(x)$ is defined by $E_{\cal R} (t)$ of (\ref{extend}),
we obtain
\beqa
W_{{\cal P}{\cal Q}} (q, N)  &=& \ld^{\frac{1}{2}(|{\cal P}| + |\cal Q|)}
\sum_U (-1)^{|U|} s_{P_+/U}(\ld^{-1} q^{-\rho}, q^{\rho}) s_{P_-/U^t}(\ld q^{\rho}, q^{-\rho}) \CR
& &~~~\sum_V (-1)^{|V|} s_{Q_+/V}(\ld^{-1} q^{- P_- -\rho} , q^{P_+ + \rho}) 
s_{Q_-/V^t}(\ld q^{P_- + \rho} , q^{- P_+ - \rho})~. \label{Hopf2}
\eeqa
Comparing (\ref{4vertex}) with (\ref{Hopf2}), we see that the identification 
$|{\cal P}| = |P_+| - |P_-|, |{\cal Q}| = |Q_+| - |Q_-|$ completes the proof of (\ref{factor}).


\vskip1cm

We would like to thank T.~Eguchi for helpful comments on the manuscript.
We also thank Y.~Konishi, S.~Matsuura, K.~Ohta, S.~Okada and S.~Minabe
for discussions.



\section*{Appendix : Normalization of Hopf Link Invariants}
\renewcommand{\theequation}{A.\arabic{equation}}\setcounter{equation}{0}
\renewcommand{\thesubsection}{A.\arabic{subsection}}\setcounter{subsection}{0}


The normalization factor of the Hopf link invariants $S_{\bullet\bullet}(q, N)$ is the partition function of
the Chern-Simons theory on $S^3$. In topological string theory it is
the contribution of constant maps to topological string amplitudes \cite{GV}. 
For $SU(N)$ theory it is given by
\beqa
S_{\bullet\bullet}(q, N) &=& \prod_{1 \leq i < j \leq N} \left( q^{\frac{1}{2}(j-i)} - q^{-\frac{1}{2}(j-i)} \right)~, \CR
                                        &=& \exp \left[ - \sum_{1 \leq i <j \leq N} \left( \frac{j-i}{2} \log q - \log (1- q^{j-i} \right)\right]~.
\eeqa
Using the strange formula for $SU(N)$
\beq
\frac{1}{2} \sum_{1 \leq i <j \leq N} (j-i)= \frac{1}{12} N (N^2 -1) = \rho_{N}^2~,
\eeq
and
\beq
\sum_{1 \leq i <j \leq N} \log (1- q^{(j-i)} ) = - \sum_{m=1}^\infty \left[  \frac{Nq^m}{m(1-q^m)} 
- \frac{q^m - q^{m(N+1)}}{m(1-q^m)^2} \right]~,
\eeq
we find
\beq
 q^{\rho_{N}^2 + \frac{N}{24}} S_{\bullet\bullet}(q, N) =M(q) \eta(q)^N N_0(q,\lambda)~,
\eeq
where $\lambda = q^N$ and 
\beq
N_0(q,\lambda) = \exp \left( - \sum_{n=1}^\infty \frac{q^n}{n(1-q^n)^2} \lambda^n \right)~,
\eeq
is regarded as non-perturbative correction to the constant map contributions. 


\end{document}